**The incredible diversity of structural and magnetic instabilities in EuTiO₃: from paramagnetic to spin glass, spin liquid and antiferromagnetic order**

Annette Bussmann-Holder[1], Reinhard K. Kremer[1], Efthymios Liarokapis[2], and Krystian Roleder[3]

[1]Max-Planck-Institute for Solid State Research, Heisenbergstr. 1, D-70569 Stuttgart, Germany

[2]Department of Physics, National Technical University of Athens, Athens 15780, Greece

[3]Institute of Physics, University of Silesia, ul. 75 Pułku Piechoty 1, 41-500 Chorzów, Poland

**Abstract** The perovskite oxide EuTiO₃ (ETO) has attracted increased scientific interest due to its potential multiferroic properties and magnetic activity above and below its structural phase transition at $T_S$=282K. Various experiments have indirectly evidenced that this transition is neither a cubic tetragonal nor the only one occurring in ETO. Here, we show new results demonstrating two further instabilities below $T_S$ based on lattice dynamics and spin-phonon interactions combined with a Landau free energy model with coupled order parameters. The new transition temperatures perfectly agree with available experimental data where further instabilities have been anticipated.

**Introduction** ETO has long been ignored experimentally and theoretically since its only interesting property was believed to be the antiferromagnetic (AFM) phase transition below $T_N$=5.5K [1]. This changed in 2001 when Katsufuji and Takagi [2] discovered a strong magneto-electric coupling in ETO, which was demonstrated by dielectric permittivity measurements. At $T_N$ the dielectric constant shows a 8% drop, which was soon after correlated with a strong transverse optic mode softening reminiscent of a ferroelectric instability [3, 4]. This finding was the starting point for increased interest in ETO since possible multiferroic properties were expected. Indeed, these were observed in thin films of ETO [5]; however, they were observed only indirectly. The analogy of ETO with SrTiO₃ (STO) was the origin to predict a structural phase transition from cubic to tetragonal around 300K, by inferring that the lattice dynamics of both systems are describable within the same model, namely the polarizability model [6 – 8], where the same parameters as of STO were also used for ETO [9]. In order to account for the AFM spin ordering, the model was complemented by the Heisenberg Hamiltonian [10] with nearest $z_1$ and second nearest neighbour $z_2$ magnetic exchange interactions $J_1$, $J_2$.

$$H = z_1 \sum_{\langle i,j \rangle} J_1 S_i S_j + z_2 \sum_{[i,j]} J_2 S_i S_j + \gamma \sum_{[i,j,n]} W_{1n}^2 S_i S_j,$$

$$H = z_1 \sum_{\langle i,j \rangle} J_1 S_i S_j + z_2 \sum_{[i,j]} (J_2 + \gamma W_{1n}^2) S_i S_j = z_1 \sum_{\langle i,j \rangle} J_1 S_i S_j + z_2 \sum_{[i,j]} (\tilde{J}_2 (T) S_i S_j). \quad (1)$$

The interaction between both Eu spins and lattice is given by a biquadratic coupling term proportional to $\gamma$ which modifies the exchange term $J_2$ as well as the lattice dynamics. $W_{1n}^2$ is the polarizability coordinate which has to be calculated self-consistently [6 – 8].

$$g_T = g_2 + 3g_4 \sum_{q,j} \frac{\hbar}{m_j \omega(q,j)} W^2(q,j) \coth \frac{\hbar \omega(q,j)}{2kT}, \quad (2)$$

for each temperature $T$, where $g_2$, $g_4$ are attractive harmonic and stabilizing anharmonic interaction constants defining $g_T$, and the summation is over all momentum $q$ and branch $j$ dependent phonon modes.

The consequence is that for temperatures $T >> T_N$ $\tilde{J}_2$ remains active through the coupling to the polarizability coordinate and adopts the T-dependence of the polarizability coordinate, whereas $J_1$ becomes meaningless above $T_N$ [10]. Importantly, the signs of $J_1$ and $J_2$ are reversed. While $J_1$ is antiferromagnetic, $J_2$ is ferromagnetic. Both are small and of the same order of magnitude [11] indicating a competition between ferro- and antiferromagnetic order and an easy switching between these states [12 – 15] . For $T > T_N$, only $\tilde{J}_2$ can trigger any magnetic order. This term is bridged by oxygen atoms along the face diagonals of the cubic perovskite lattice and forms two interpenetrating sign reversed ferromagnetic tetrahedra which are magnetically compensated in the cubic structure. Since they are coupled to the structure, the lattice dynamics of ETO are modified, which is reflected in an enhancement of the structural phase transition temperature and carries transient magnetic properties [16].

**Theory** In spite of the intriguing similarities between ETO and STO, namely softening of a transverse optic and a zone boundary acoustic mode, same lattice parameters, comparable ionic radii of $Sr^{2+}$ and $Eu^{2+}$ together with the $d^0$ configuration of $Ti^{4+}$, their double well potentials in the polarizability coordinate (equ. 2) differ considerably caused by the spin-lattice coupling [11]. While the one of STO is broad and shallow, the one of ETO is narrow and deep (Figure 1). In addition, the phase transition temperature of ETO can be influenced by a magnetic field without observing any global magnetic order [16, 17]. This missing order became the topic of a series of muon spin rotation (μSR) experiments, which demonstrated spin activity below and above $T_S$ but without providing its origin [17 – 20] in terms of ferro- or antiferromagnetism. Similarly, birefringence $\Delta n$ measurements could be influenced by a magnetic field by shifting its onset by several degrees in temperature in a small applied magnetic field [20, 21]. Another outcome of these measurements was the observation of a new temperature scale since below $T_S$ perfect Landau type increase in birefringence was reported, however, with an abrupt change in slope around $T^* \approx 210K$. A further change in $\Delta n$ seems to occur well below $T^*$ around 100K, but the experimental data are insufficient to confirm this. These novel temperature scales were indirectly confirmed by μSR and dielectric results [22] but could not be seen in high-resolution x-ray diffraction data, which were limited due to experimental temperature accessibility [23, 24]. A puzzle appears in Raman scattering experiments where it was assumed that modes active in STO would also be detectable in ETO. Instead, nothing was observed [25, 26].

Here, we give an amazing theoretical explanation that fully explains the experimental results by combining the self-consistently derived lattice dynamics with a Landau free energy expansion with coupled order parameters valid for a second-order phase transition [27 – 29]. Is is in the spirit of the book by Toledano and Toledano [30] and reminiscent of related approaches [31 – 35] where specific features of Landau theory have been addressed, namely the ordered phase only [31], the relations to renormalization group theory [32], coupled order parameters [33, 34] or symmetry relation [35]. As compared to these examples our approach is less sophisticated since primary symmetry aspects are not included and renormalization or statistical mechanics and topology are neglected. This is definitely a shortcome but provides more transparency, diminishes the parameter space and enables easy solvability. We choose as order parameters the polarizability coordinate $W$ since this is the most relevant quantity for the lattice dynamics of perovskite oxides in general, specifically STO and ETO. The other order parameter refers to the hidden magnetism stemming from the hybridization of the spins and the lattice which we define by $M$. The background of this choice is based on the fact that the difference between the double-well potentials of ETO and STO must be related to the inherent magnetic properties of ETO. The raw, uncoupled potentials, and free energies, respectively, as shown in Fig. 1, are defined by the harmonic attractive and anharmonic fourth-order coupling constants $g_2$ and $g_4$, which have been derived self-consistently within the polarizability model [11].

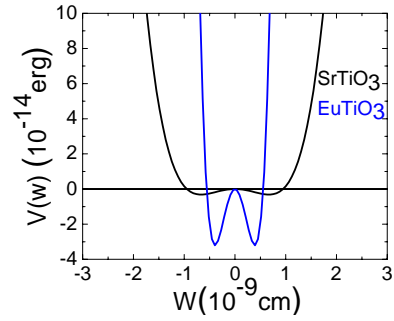

**Fig. 1** Free energies of STO (black) and ETO (blue) as a function of the polarizability coordinate [11].

Without the features of the "hidden" magnetism appearing in ETO, its potential should be identical to the one of STO. However, the mentioned differences are ascribed to the presence of the $4f^7$ spins of $Eu^{2+}$.

Correspondingly, the Landau free energy expansion of the two order parameters adopts the following form:

$$F(P, M) = \frac{1}{2}\alpha W^2 + \frac{1}{4}\beta W^4 - \frac{1}{2}\gamma W^2 M^2 + \frac{1}{2}cM^2 + \frac{1}{4}dM^4 \qquad (3)$$

Identical to the lattice dynamical Hamiltonian used for ETO a biquadratic coupling proportional to $\gamma$ is introduced. We follow the standard procedure of Holakovsky [27] and minimize equ. 3 with respect to $M$ which yields,

$$\frac{\partial F}{\partial M} = M(-\gamma W^2 + c + dM^2) = 0 \qquad (4)$$

and

$$\frac{\partial^2 F}{\partial M^2} = (-\gamma W^2 + c + dM^2) + 2dM^2 > 0 \qquad (5)$$

$M = 0$ corresponds to the case of a single order parameter, i.e., $\gamma$=0. The solution with $M$, $\gamma \neq 0$ satisfies the inequality (5) and yields:

$$M^2 = \frac{\gamma W^2 - c}{d} \qquad (6)$$

Substituting this value of $M^2$ into equ. (3) the following function for the free energy is obtained:

$$F(W) = -\frac{c^2}{4d} + \frac{W^2}{2}\left(\alpha + \frac{\gamma c}{d}\right) + \frac{W^4}{4}\left(\beta - \frac{\gamma^2}{d}\right) \tag{7}$$

For a second order phase transition to occur the coefficient of $W^4$ must be positive requiring that

$$d\beta > \gamma^2 \tag{8}$$

The parameters for $\alpha$, $\beta$, $c$, $d$ are those derived within the self-consistent phonon approximation (SPA), namely, $\alpha = g_2(\text{STO}) = -1.41$, $\beta = g_4(\text{STO}) = 1.57$, $c = g_2(\text{ETO}) = -41.3806$, $d = g_4(\text{ETO}) = 133.5556$, which implies that $\gamma < 14.48$. The temperature dependence of $F(W)$ is introduced in a Landau typical way, with $\alpha = g_{2(\text{STO})}(T - T_S): T > T_{S,}$, $\alpha = -2g_{2(\text{STO})}(T - T_S): T < T_S$; $c = g_{2(\text{ETO})}(T - T_M): T > T_{M,}$, $\alpha = -2g_{2(\text{ETO})}(T - T_M): T < T_M$. Note, that the above given values for the double-well defining quantities have been determined self-consistently and are valid for all temperatures. Their determination is thus not dependent on a specific temperature where they have been evaluated numerically, but the specifics of the SPA are that their values need to be the same for any temperature. It is also important to mention that the measurable quantities, the spontaneous polarization $P_S$ and the spontaneous magnetization $M_S$ should not be mixed up with the order parameters, i.e. the spontaneous polarization and the spontaneous magnetization are zero above $T_S$, $T_M$, respectively whereas the order parameter is zero only at $T = T_S$, $T = T_M$ and finite above and below the corresponding transition temperature.

We know from experiments that for STO $T_S = 105$ K and for ETO $T_S = T_M = 282$ K. The free energy equ. 7 is a function of temperature and $\gamma$. This dependence is plotted as a function of $W$, which has the dimensions of a length scale for various values of $\gamma$ in Figs. 2.

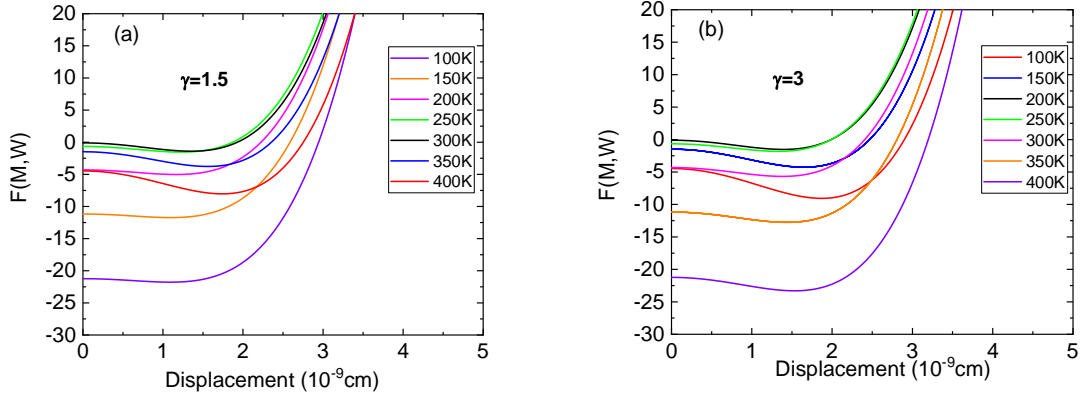

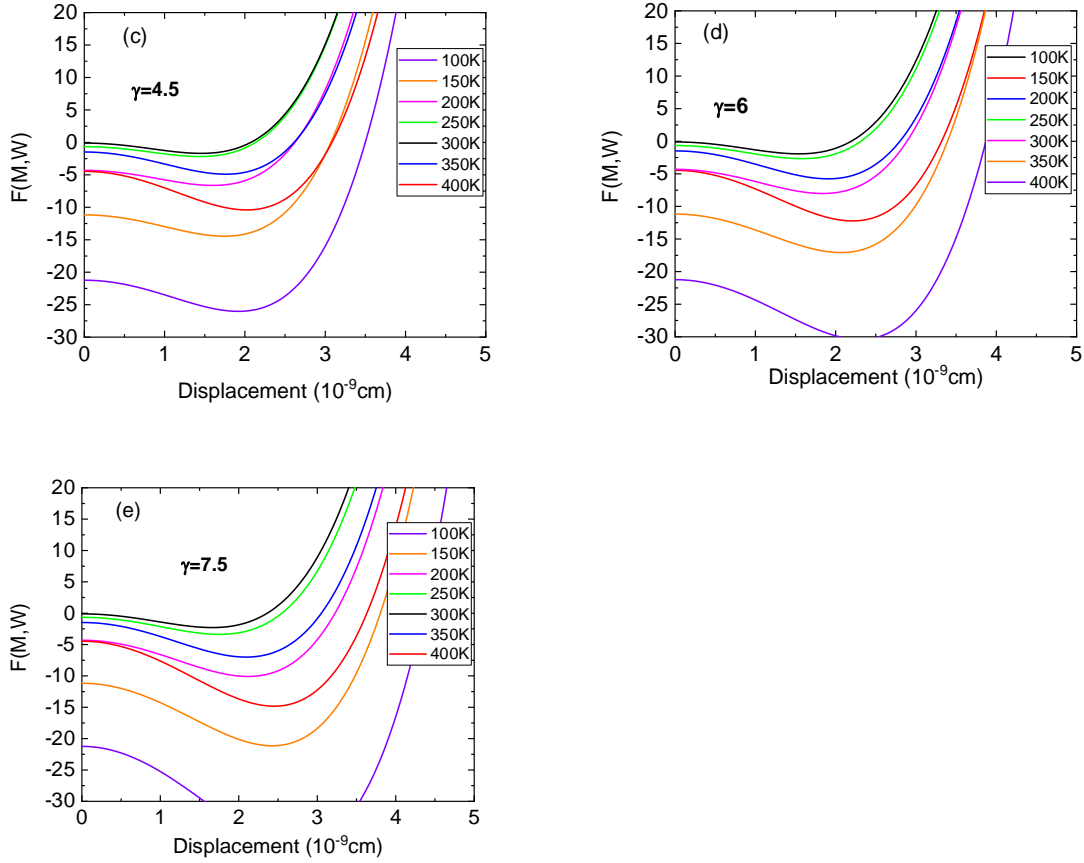

**Figs. 2** Free energy as a function of temperature and the order parameters *W, M* which are given in terms of a length scale, namely the displacement, in order to account for the correct dimensions. From left up to right down the coupling parameter *γ* increases from (a) 1.5 to (b) 3, (c) 4.5, (d) 6, (e) 7.5 as given in the Figures.

What is apparent at first glance in Figs. 2a-e, is that *γ* leads to a broadening of the potential for all temperatures. Furthermore, the potential depth is increased with increasing *γ*. The temperature variation has the effect that for $T < T_S$, the potential is broadest and deepest, i.e., in order to overcome the barrier, the largest energy is required. Once $T > T_S$ of STO (≈105K) is reached, its depth and spread reduce systematically until $T_S$=282K of ETO is reached where an almost flat potential appears since *M* dominates from there on. For $T > T_S$ the double-well character reappears with increasing temperature. From then on this gets deeper and broader but never adopts the broadness and depth as observed for $T < T_S$ (STO). A better understanding of this development can be achieved by inspecting the temperature and *γ* dependence of the squared order parameters $W^2$ and $M^2$ which are shown in Figs. 3.

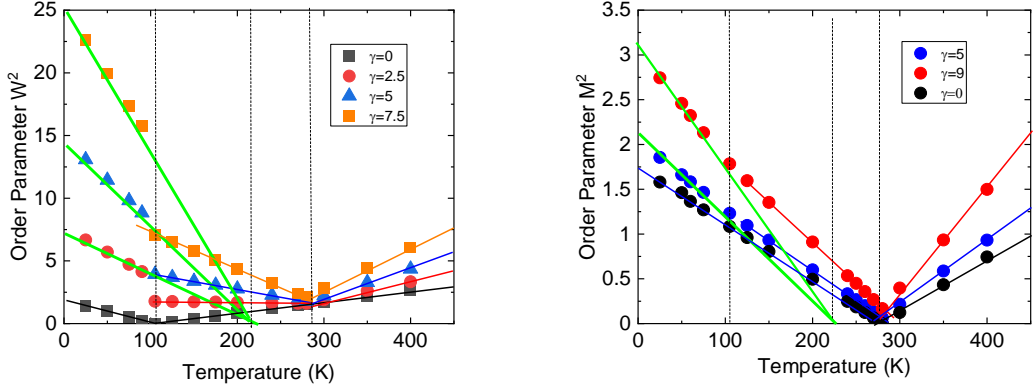

**Figs. 3** Temperature dependence of the order parameters $M^2$ (right), $W^2$ (left), for various values of $\gamma$.

The most striking observation in the two figures is the fact that $W^2$ is almost an order of magnitude larger than $M^2$ in accordance with the "hidden" character of any magnetization. In both cases, however, $\gamma$ substantially increases both order parameters as compared to the uncoupled ($\gamma$=0) case (black squares and dots). Upon fitting the order parameters by a linear in $T$ behaviors as required by Landau theory, changes in slope are visible which can be extrapolated to zero around $T^* = 210$K for both order parameters. This is a compelling evidence for another phase transition to occur in ETO. Most interestingly, this transition temperature is independent of $\gamma$. The dependence of the polarizability order parameter on temperature and $\gamma$ shows besides of $T^*$ a discontinuous break in its temperature evolution at $T_S \approx 105$K which corresponds approximately to the structural phase transition temperature of STO and reflects the close similarity between both compounds. The observed discontinuity around 105K is very suggestive to conclude that the transition is of first order.

**Experiment** Since, as emphasized above, the only difference between STO and ETO lies in the presence of largely localized $4f^7$ states of $Eu^{2+}$ all observed transitions can be associated with some kind of magnetism. This can be easiest tested experimentally by probing characteristic properties in a small magnetic field. As a first test the magnetic susceptibility times temperature, which should be independent of $T$ for $T >>> T_N$ is reinvestigated (Fig. 4) over a much broader temperature region as compared to Ref. 36 where these data are shown in the inset to Fig. 4.

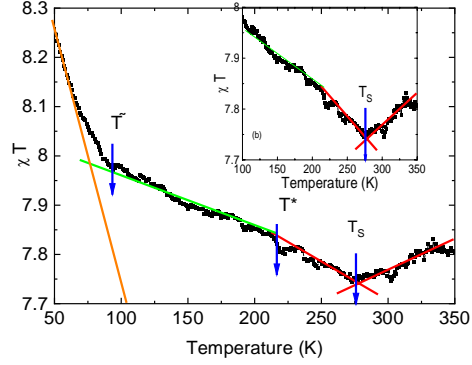

**Fig. 4** Magnetic susceptibility times temperature χT as a function of *T*. The inset shows the results of Ref. 28 where the data have been plotted for a limited temperature region to highlight the anomalies of χT around $T_S$.

On the enlarged temperature scale of Fig. 4 all predicted temperatures are visible where *T\** appears as a change in slope and a sharp increase of χT arises at $\bar{T} \approx 105$K. . A further experiment is the birefringence of ETO reported previously [36 – 38], where several deviations of the typical Landau behavior in it showed obvious deviations at $T_S$, *T\** and an undefined lower temperature, which we call $\bar{T}$. This temperature was observed when continuing the measurements to lower temperatures, as shown in Fig. 5. Because the sample had previously been exposed to a small magnetic field, all transitions are shifted to lower T. This field effect remained and is well visible in Fig. 5. Again, all temperature scales appear.

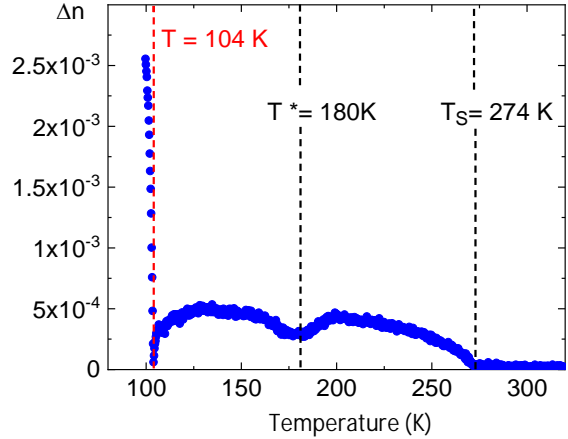

**Fig. 5** Birefringence results for an ETO film on a STO substrate. Three different phase transitions are clearly seen, which are shifted to slightly lower temperatures as compared to magnetic susceptibility data caused by the memory effect of previously applying magnetic fields of induction B<0.1T to the film. A drastic change of the birefringence below 104 K proves the existence of the third transition near $\bar{T}$ in Fig. 4.

Since we have shown that all phases below $T_S$ (ETO) exhibit intrinsic "ferro"magnetic properties, a symmetry lowering at 282K lifts the degeneracy of the interpenetrating ferromagnetic tetraedra whereby true ferromagnetic domain formation is enabled which, however, are not linked to macroscopic magnetism. Upon inspecting earlier magnetic susceptibility data [36] again, it is very suggestive to identify this transition with a spin liquid formation one [39, 40] which opposite to known ones is triggered by the lattice since it follows the same temperature dependence as the soft zone boundary transverse acoustic mode. Far above $T_S$ ETO has a magnetically neutral cubic structure consisting of two interpenetrating energetically degenerate tetraedra with sign reversed spin states. At $T_S$ a cubic to tetragonal phase transition is expected analogous to the one observed in STO and other perovskite oxides. However, instead a transition to much lower symmetry takes place which lifts the degeneracy of the sublattices and introduces frustration suggesting that within the ground states there are local fluctuations which take place independently in different parts of a large sample. Experimentally this has been observed in the tiny kink in the inverse magnetic susceptibility (Fig. 4, inset) and through µSR experiments [20, 21]. Such a phase is known to break time reversal symmetry and provides an intriguing explanation for the missing Raman signals. Consequently, at the following transitions at 210K and 100K enhanced magnetic activity sets in caused by further symmetry lowering. We speculate that the phase between 210K and 100K might be best described as magnetic polarons [41] or as a spin liquid as early suggested in [41 – 43] which is frozen below 100K to finally adopt AFM order at very low temperatures. It is important to note, that $T_N$ is very small and positive, $J_1$, $J_2$ are of the same order of magnitude with opposite signs signaling magnetic frustration.

In the above analysis anisotropy of γ has not been taken into account in order to keep the number of parameters as small as possible and for the sake of transparency. This has, however, the disadvantage that conclusions about the space groups associated with the symmetry lowering at $T_S$, $T^*$ and $\tilde{T}$ are not possible. By including anisotropy in γ and following the procedure of Thomas and Müller [44] the free energy adopts a new form as described in the Supplementary Material section which allows to conclude about possible symmetry lowering associated with the different phases. In addition, it enables to analyse the domain structure as observed by birefringence.

It is important to note, that neither first principles calcualtions [45 – 47] nor densitiy functional theory [48 – 52] has neither predicted the 282K phase transition nor other instabilities in ETO, but has shown that the low temperature AFM phase is energetically only a few millivolts lower in energy as compared to other spin ordered phases.

**Conclusions** In conclusion, we predict for ETO all together two more phase transitions to occur below $T_S$=282K, one at $T^*$ = 210K and another one near $\tilde{T}$ ≈105K. These transitions have been obtained within Landau theory of coupled order parameters combined with the polarizability model, where its self-consistently derived coupling parameters have been taken as input for the Landau free energy expansion parameters. It is amazing how predictive this methodology is and how well it describes experimental findings, which provide transient and almost hidden new properties. These new transitions are a challenge for further experiments, and we suggest investigating them with an applied magnetic field. In addition, we associate them with spin glass and spin liquid states with rather high temperatures driven by their coupling to the lattice.

**Supplementary Material**

In the paper by Thomas and Müller [44] a model Hamiltonian for the study of phase transitions in perovskite compounds $ABO_3$ involving rotations of $BO_6$ octahedra is studied. Depending on the relative magnitude of the anharmonic coefficients, transitions to the tetragonal or to the trigonal phase are obtained. This model is extended to include a coupling with a second order parameter yielding a Landau free energy of the form:

$$F(P,M) = \frac{1}{2}\alpha\left(W_x^2 + W_y^2 + W_z^2\right) + \frac{1}{4}\beta\left(W_x^4 + W_y^4 + W_z^4\right) + \frac{e}{2}\left(W_x^2 W_y^2 + W_y^2 W_z^2 + W_z^2 W_x^2\right) - \frac{1}{2}\left(\gamma_1 W_x^2 + \gamma_2 W_y^2 + \gamma_3 W_z^2\right)M^2 + \frac{1}{2}cM^2 + \frac{1}{4}dM^4 \tag{1A}$$

Identical to the lattice dynamical Hamiltonian used for ETO a biquadratic coupling proportional to $\gamma$ is introduced, which in general will be differ for the three space directions. Following the standard procedure of Holakovsky [25] and minimizing the free energy with respect to $M$ yields:

$$\frac{\partial F}{\partial M} = M\left(-\gamma_1 W_x^2 - \gamma_2 W_y^2 - \gamma_3 W_z^2 + c + dM^2\right) = 0 \tag{2A}$$

and

$$\frac{\partial^2 F}{\partial M^2} = \left(-\gamma_1 W_x^2 - \gamma_2 W_y^2 - \gamma_3 W_z^2 + c + dM^2\right) + 2dM^2 > 0 \tag{3A}$$

where $M = 0$ corresponds to the case of SrTiO$_3$. The other solution with $M \neq 0$, satisfies the inequality equ. 3A and gives:

$$M^2 = \frac{\gamma_1 W_x^2 + \gamma_2 W_y^2 + \gamma_3 W_z^2 - c}{d} \tag{4A}$$

Since $c = c_0(T - T_M)$ $M^2$ is always positive for any temperature $T_M > T$. Substituting this value of $M^2$ the following function for the free energy is obtained,

$$F(W) = \frac{1}{2}\alpha\left(W_x^2 + W_y^2 + W_z^2\right) + \frac{1}{4}\beta\left(W_x^4 + W_y^4 + W_z^4\right) + \frac{e}{2}\left(W_x^2 W_y^2 + W_y^2 W_z^2 + W_z^2 W_x^2\right) - \frac{\left[\gamma_1 W_x^2 + \gamma_2 W_y^2 + \gamma_3 W_z^2 - c\right]^2}{4d} \tag{5A}$$

This can be rewritten as:

$$F(W) = -\frac{c^2}{4d} + \frac{W_x^2}{2}\left(\alpha + \frac{\gamma_1 c}{d}\right) + \frac{W_y^2}{2}\left(\alpha + \frac{\gamma_2 c}{d}\right) + \frac{W_z^2}{2}\left(\alpha + \frac{\gamma_3 c}{d}\right) + \frac{W_x^4}{4}\left(\beta - \frac{\gamma_1^2}{d}\right) + \frac{W_y^4}{4}\left(\beta - \frac{\gamma_2^2}{d}\right) + \frac{W_z^4}{4}\left(\beta - \frac{\gamma_3^2}{d}\right) + \frac{e}{2}\left(W_x^2 W_y^2 + W_y^2 W_z^2 + W_z^2 W_x^2\right) \tag{6A}$$

For a second order phase transition to occur the coefficient of $W^4$ must be positive implying that

$$d\beta > \gamma_i^2 \text{ (i=1, 2, 3)} \tag{7A}$$

The parameters $\alpha, \beta, c, d$ are identical to those derived in the main text. Due to the anisotropy of $\gamma$ new transition temperatures can be defined in the case of three different values $\gamma$. All of them are obtained from the function of the leading order in $W^2$ which depends on temperature,

$$T_c(\gamma) \equiv \frac{a_0 dT_S + \gamma c_0 T_M}{a_0 d + \gamma c_0} \tag{8A}$$

If $\gamma_2 = \gamma_3 = 0$, the results are identical to the case as described in the main text. For the more complicated cases $\gamma_1 \neq \gamma_2 \neq 0, \gamma_3 = 0$ and recursions in the index and also all three $\gamma_i$ (i=1, 2, 3) being different and unequal from zero, three new transition temperatures appear which are functions of $\gamma_i$. By using predefined values of $\gamma_i$ the symmetry of the low temperature phases can be obtained, however, such a procedure has ambuiguity and needs to be justified by experimental observations. Another option is to use the relations among the different $W_i^2$ terms in order to enable the identification of the low temperature symmetry groups. Here an option would be to derive these from high resolution X-ray diffraction data which poses however a challenge due to the really tiny differences in the involved length scales.